\documentclass[letterpaper,twocolumn,10pt]{article}
\usepackage{usenix}
\usepackage{subfig}
\usepackage{graphicx,amsmath,amssymb,xspace}
\usepackage{color}
\usepackage{tabularx}
\usepackage{cite}
\usepackage{import}

\newcommand{\eg}{e.g.,\xspace}

\newcommand{\etc}{etc.\@\xspace}
\newcommand{\ie}{i.e.,\xspace}

\newcommand{\unit}[1]{\ensuremath{\, \mathrm{#1}}}
\newcommand{\rxname}{SPREE\xspace}
\graphicspath{{./figures/}}

\begin{document}

\date{}
\title{\rxname: A Spoofing Resistant GPS Receiver}

\author{\rm Aanjhan Ranganathan, Hildur \'{O}lafsd\'ottir, Srdjan Capkun\\
Department of Computer Science\\
ETH Zurich, Switzerland\\
} 

\maketitle


\subsection*{Abstract} 
Global Positioning System (GPS) is used ubiquitously in a wide variety of applications ranging from navigation and tracking to modern smart grids and communication networks. However, it has been demonstrated that modern GPS receivers are vulnerable to signal spoofing attacks. For example, today it is possible to change the course of a ship or force a drone to land in an hostile area by simply spoofing GPS signals. Several countermeasures have been proposed in the past to detect GPS spoofing attacks. These countermeasures offer protection only against naive attackers. They are incapable of detecting strong attackers such as those capable of seamlessly taking over a GPS receiver, which is currently receiving legitimate satellite signals, and spoofing them to an arbitrary location. Also, there is no hardware platform that can be used to compare and evaluate the effectiveness of existing countermeasures in real-world scenarios. In this work, we present \rxname, which is, to the best of our knowledge, the first GPS receiver capable of detecting all spoofing attacks described in literature. Our novel spoofing detection technique called auxiliary peak tracking enables detection of even a strong attacker capable of executing the seamless takeover attack. We implement and evaluate our receiver against three different sets of GPS signal traces and show that \rxname constrains even a strong attacker (capable of seamless takeover attack) from spoofing the receiver to a location not more than 1 km away from its true location. This is a significant improvement over modern GPS receivers that can be spoofed to any arbitrary location. Finally, we release our implementation and datasets to the community for further research and development.

\section{Introduction}
Today, a number of security- and safety-critical applications rely on Global Positioning Systems (GPS)~\cite{misra2006global} for positioning and navigation. A wide-range of applications such as civilian and military navigation, people and asset tracking, emergency rescue and support, mining and exploration, atmospheric studies, smart grids, modern communication systems use GPS for localization and timing. GPS is a satellite based navigation system that consists of more than 24 satellites orbiting at more than 20,000 km above the earth. Each satellite continuously broadcasts data called ``navigation messages'' containing its precise time of transmission and the satellite’s location. The GPS receiver on the ground receives each of the navigation messages and estimates their time of arrival.  Based on the time of transmission that is contained in the navigation message itself and its time of arrival, the receiver computes its distance to each of the visible satellites. Once the receiver acquires the navigation messages from at-least four satellites, the GPS receiver estimates its own location and precise time using the standard technique of multilateration.

However, the civilian GPS navigation messages that are transmitted by the satellites lack any form of signal authentication. This is one of the prime
reasons GPS is vulnerable to \emph{signal spoofing} attacks. In a GPS
spoofing attack, an attacker transmits specially crafted signals identical to
those of the satellites but at a higher power that is sufficient enough to overshadow the legitimate satellite signals. The GPS receiver then computes a false location and time based on the stronger spoofing signal transmitted by the attacker. As a result, today, it is possible to spoof a GPS receiver to any arbitrary location. For example, researchers have demonstrated the insecurity of GPS-based navigation by diverting the course of a yacht using spoofed GPS signals~\cite{yacht_spoofing}. A similar hijack was also successfully executed on a drone using a GPS spoofer that costs less than \$1000. More recently, researchers demonstrated a GPS signal generator that can be built for less than \$300~\cite{lowcost_gps}. The increasing availability of low-cost radio hardware platforms~\cite{ettus_research} make it feasible to execute such attacks with less than few hundred dollars worth of hardware equipment. More advanced attacks were demonstrated in~\cite{NighswanderCCS2012,Tippenhauer2011} in which the attackers \textit{takeover} a target receiver that is already locked (continuously receiving navigation messages) on to authentic satellite signals without the receiver noticing any disruption or loss of navigation data. It was shown that a variety of commercial GPS receivers were vulnerable and in some cases even caused permanent damage to the receivers. It is thus evident that these threats are real and it is important to secure GPS from such signal spoofing attacks.

Although spoofing attacks can be, to a certain extent, mitigated by adding cryptographic authentication to the navigation messages (e.g., military GPS systems where the spreading codes are secret), their use requires distribution and management of shared secrets, which makes them impractical for majority of applications. Even with cryptographic authentication, the system is not protected against relay attacks where an attacker simply records and replays the radio signals to the receiver~\cite{Papadimitratos2008}. Several countermeasures that did not require cryptographic authentication were proposed in recent years either to detect or to mitigate signal spoofing attacks. They rely on detecting anomalies in certain physical characteristics of the signal such as received satellite signal strength, ambient noise floor levels, automatic gain control~\cite{Akos2012} values and other data that are readily available as \textit{receiver observables} on modern GPS receivers. Some other countermeasures leveraged the signal's spatial characteristics~\cite{Montgomery2009,psiaki2013gnss} such as the received GPS signal's direction or angle of arrival. All the above mentioned countermeasures are ineffective against attackers capable of manipulating navigation message contents in real time or a seamless takeover attack~\cite{NighswanderCCS2012,Tippenhauer2011}. Additionally, majority of these solutions are not reliable in an environment with strong multipath (signal copies that reach the receiver with a time delay due to reflections in the environment \etc) or in the case of a mobile receiver. Moreover, today there is no receiver platform that can be used to compare and evaluate the effectiveness of these countermeasures in real-world scenarios.   

In this work, we present a novel GPS receiver which we refer to as \rxname and make the following contributions: \rxname is to the best of our knowledge, the first commercially off the shelf, single-antenna, receiver capable of detecting or significantly limiting all known GPS spoofing attacks described in literature. \rxname does not rely on GPS signal authentication and can therefore be used to detect both civilian and military GPS spoofing attacks. Additionally, it is designed to be standalone and does not depend on other hardware such as antennas, additional sensors or alternative sources of location information (like maps or inertial navigation systems). In \rxname, we introduce a novel spoofing detection technique called auxiliary peak tracking that limits even a strong attacker (\eg seamless takeover) from being able to move (spoof) a receiver to any arbitrary location or time. We leverage the presence of authentic signals in addition to the attacker's signals to detect spoofing attacks. We implement \rxname by modifying an open source software-defined GPS receiver~\cite{GNSS-SDR11} and evaluate it against different signal data sets including the de-facto standard of publicly available repository of GPS signal spoofing traces (Texas Spoofing Battery (TEXBAT)~\cite{humphreys2012texas}). Furthermore, we evaluate \rxname against COTS GPS simulators and our own traces obtained through an extensive wardriving effort of over 200 km. Our analysis shows that \rxname can reliably detect any manipulations to the navigation message contents. In addition, \rxname severely limits even strong attackers capable of taking over a receiver that is currently locked (receiving and decoding) on to legitimate satellite signals without being noticed. Our evaluations showed that such a strong attacker could offset the \rxname's location to a maximum of 1 km away from its true location. Finally, we release our implementation and a set of recorded GPS signal traces used for evaluating \rxname to the community for further research and development.

\section{GPS Overview}
\label{sec:gps-overview}

\subsection{GPS Satellite System}

GPS comprises more than 24 satellites orbiting the earth at more than 20,000~km above the ground. Each satellite is equipped with high-precision atomic clocks and hence the timing information available across all the satellites are in
near-perfect synchronization. Each satellite transmits messages referred to as the \textit{navigation messages} that are spread using pseudorandom codes that are unique to a specific satellite. 

The navigation data transmitted by each of the satellites consists of a $1500$ bit long data frame which is divided into 5 subframes~\cite{borre2007software}. Subframes 1, 2 and 3 carry the same data across each frame. The data contained in subframes 4 and 5 is split into 25 pages and is transmitted over 25 navigation data frames. The navigation data is transmitted at $50\unit{bps}$ with the duration of each subframe being 6 seconds. Each frame lasts 30 seconds and the entire navigation message, containing 25 such frames, takes 12.5 minutes to be received completely by a receiver. The first subframe mainly contains satellite clock information. The second and third subframes contain the ephemeris \ie information related to the satellite's orbit and is used in computing the satellite position. Subframes 4 and 5 contain the almanac data \ie the satellite orbital and clock information with reduced precision for all satellites. 


\subsection{GPS Receiver}
\label{sub:gps-rx-overview}
\begin{figure}[t]
  \centering
  \includegraphics[width=0.9\columnwidth]{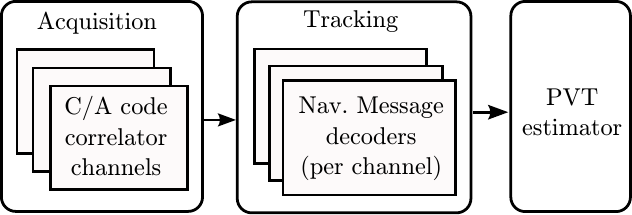}
  \caption{GPS receiver architecture. The RF front-end (not shown in Fig) preprocesses the received satellite signals. The acquisition module searches for any visible satellite signals and if detected forwards it to the tracking module. The tracking module decodes the navigation data which is used in estimating the position, velocity and time.}
  \label{fig:gps-rx-architecture}
\end{figure}

A typical GPS receiver consists of four main building blocks: (i) RF front-end, (ii) Acquisition module, (iii) Tracking module and (iv) Position, Velocity, Time (PVT) estimator module. The RF front-end block pre-processes the received satellite signals and forwards it to the acquisition module. The acquisition module is responsible for searching for any satellite signals and forwarding the signal to the tracking module when a visible satellite signal is detected. The tracking module decodes and extracts the navigation data from the acquired signal and sends it to the PVT estimator module for computing the receiver’s location and time.

\begin{figure}[t]
  \includegraphics[width=0.95\columnwidth]{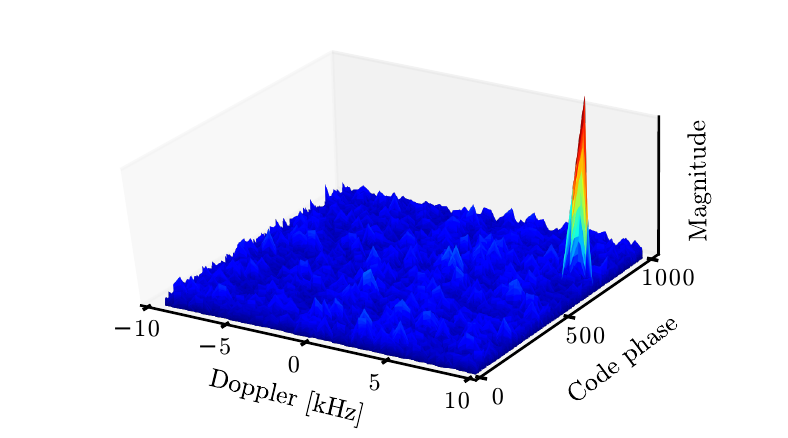}
  \caption{GPS Signal Acquisition. The result of the correlation for a real satellite signal acquisition.}
  \label{fig:acquisition}
\end{figure}

The acquisition module searches for satellite signals by correlating its own replica of the pseudorandom code corresponding to each of the satellites. In addition, the carrier frequency of the satellite signals can differ from its true value due to the relative motion of the satellite and the receiver itself (doppler effect). Thus, in order to detect any visible satellite signal, the receiver performs a two-dimensional search. First, it has to search through all possible delays (phase) of the pseudorandom code. Second, the receiver must account for frequency errors that occur due to the doppler effect and other environmental interferences. Figure~\ref{fig:acquisition} shows the output of a signal acquisition phase. If the code and doppler searches result in a peak above a certain threshold the GPS receiver then switches to tracking and demodulating the navigation message data. The decoded data is used to estimate the receiver's range or distance from each of the visible satellites. In-order to determine the range, the receiver needs the satellite signal's transmission and reception time. The transmission time of each subframe is found in the navigation message and the reception time is estimated by the receiver. It is important to note that the satellite clocks are in tight synchronization with each other while the receiver's clock (not using atomic crystals) contain errors and biases. Due to the receiver's clock bias, the estimated ranges are referred to as \emph{pseudoranges}. The receiver requires at-least four pseudoranges to estimate its position after eliminating the effect of receiver clock bias.

\section{GPS Spoofing Attacks}
\label{sec:gps-spoofing-attack}

A GPS signal spoofing attack is a physical-layer attack in which an attacker transmits specially crafted radio signals that are identical to authentic satellite signals. Civilian GPS is easily vulnerable to signal spoofing attacks. This is due to the lack of any signal authentication and the publicly known spreading codes for each satellite, modulation schemes and data structure. In a signal spoofing attack, the objective of an attacker may be to force a target receiver to (i) compute a false geographic location, (ii) compute a false time or (iii) disrupt the receiver by transmitting unexpected data. Due to the low power of the legitimate satellite signal at the receiver, the attacker's spoofing signals can trivially overshadow the authentic signals. During a spoofing attack, the GPS receiver locks (acquires and tracks) on to the stronger signal \ie the attacker's signals, ignoring the legitimate satellite signals. This results in the receiver computing a false position, velocity and time based on the spoofing signals. 

An attacker can influence the receiver's position and time estimate in two ways: (i) manipulating the contents of the navigation messages (\eg location of satellites, navigation message transmission time) and/or (ii) modify the arrival time of the navigation messages. The attacker can manipulate the arriving time by temporally shifting the navigation message signals while transmitting the spoofing signals. We classify the different types of spoofing attacks based on how synchronous (in time) and consistent (with respect to the contents of the navigation messages) the spoofing signals are in comparison to the legitimate GPS signals currently being received at the receiver's true location.\\ 
\begin{figure}[t]
  \centering
  \includegraphics[width=0.8\columnwidth]{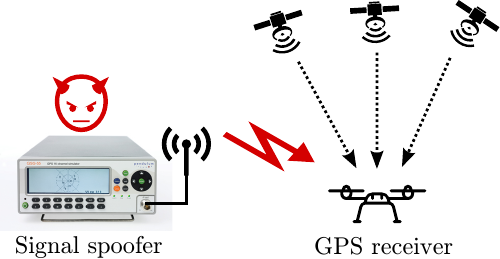}
  \caption{Spoofing attack. The attacker uses a commercial GPS simulator to transmit signals identical to legitimate satellite signals but with a higher power to overshadow the legitimate signals. The receiver computes a false location and time based on the spoofing signals.}
  \label{fig:spoofing-attack}
\end{figure}

\noindent\textit{Non-Coherent and Modified Message Contents:} In this type of a attack, the attacker's signals are both unsynchronized and contain different navigation message data in comparison to the authentic signals. Attackers who use GPS signal generators~\cite{spectracom,labsat} to execute the spoofing attack typically fall under this category. An attacker with little know-how can execute a spoofing attack using these simulators due to their low complexity, portability and ease of use. Some advanced GPS signal generators are even capable of recording and replaying signals, however not in real-time. In other words, the attacker uses the simulator to record at one particular time in a given location and later replays it. Since they are replayed at a later time, the attacker's signals are not coherent and contain different navigation message data than the legitimate signals currently being received.\\

\noindent\textit{Non-Coherent but Unmodified Message Contents:} In this type of attack, the navigation message contents of the transmitted spoofing signals are identical to the legitimate GPS signals currently being received. However, the attacker temporally shifts the spoofing signal thereby manipulating the spoofing signal's time of arrival at the target receiver. For example, attackers capable of real-time record and replay of GPS signals fall under this category as they will have the same navigation contents as that of the legitimate GPS signals, however shifted in time. The location or time offset caused by such an attack on the target receiver depends on the time delay introduced both by the attacker and due to the propagation time of the relayed signal. The attacker can precompute these delays and successfully spoof a receiver to a desired location.\\

\noindent\textit{Coherent but Modified Message Contents:} The attacker generates spoofing signals that are synchronized to the authentic GPS signals. However, the contents of the navigation messages are not the same as that of the currently seen authentic signals. For example, attacks such as those proposed in~\cite{NighswanderCCS2012} can be classified under this category. Nighswander et al.~\cite{NighswanderCCS2012} present a Phase-Coherent Signal Synthesizer (PCSS) that is capable of generating a spoofing signal with the same code phase as the legitimate GPS signal that the target receiver is currently locked on to. Additionally, the attacker modifies the contents of the navigation message in real-time (and with minimal delay) and replays it to the target receiver. A variety of commercial GPS receivers were shown to be vulnerable to this attack and in some cases, it even caused permanent damage to the receivers.\\


\noindent\textit{Coherent and Unmodified Message Contents:} Here, the attacker does not modify the contents of the navigation message and is completely synchronized to the authentic GPS signals. Even though the receiver locks on to the attacker's spoofing signals (due to the higher power), there is no change in the location or time computed by the target receiver. Therefore, this is not an attack in itself but is an important first step in executing the seamless takeover attack. \\
\begin{figure*}[t]
  \centering
  \includegraphics[width=0.80\textwidth]{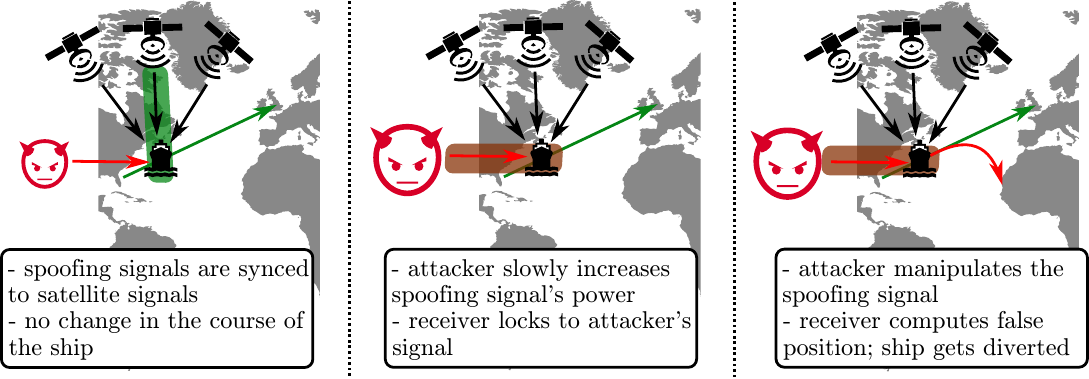}
  \caption{Seamless takeover attack. The receiver is locked on to the legitimate satellite signals. The spoofing signal is synchronized with the legitimate signal and has the same navigation contents. Next, the attacker slowly increases the power of the spoofing signal. The receiver stops tracking the legitimate signals and locks on to the attacker's signal. Finally, the attacker temporally shifts the spoofing signal causing the receiver to compute a false location and thereby changing the ship's route.}
  \label{fig:seamless-takeover}
\end{figure*}

\subsection{Seamless Takeover Attack}
The seamless takeover attack is considered one of the strongest attacks in literature. In a majority of applications, the target receiver is already locked on to the legitimate GPS satellite signals. The goal of an attacker is to force the receiver to stop tracking the authentic GPS signals and lock on to the spoofing signals without causing any signal disruption or data loss. This is because the target receiver can potentially detect the attack based on the abrupt loss of GPS signal. Consider the example of a ship on its way from the USA to the UK as shown in Figure~\ref{fig:seamless-takeover}. The GPS receiver on the ship is currently locked on to the legitimate satellite signals. In a seamless takeover attack, first, the attacker transmits spoofing signals that are synchronized with the legitimate satellite signals and are at a power level lower than the received satellite signals. The receiver is still locked on to legitimate satellite signals due to the higher power and hence there is no change in the ship's route. The attacker then gradually increases the power of the spoofing signals until the target receiver stops tracking the authentic signal and locks on to the attacker's spoofing signals. Note that during this takeover, the receiver does not see any loss of lock, in other words, the takeover was seamless. Even though the target receiver is now locked on to the attacker, there is still no change in the route as the spoofing signals are both coherent with the legitimate satellite signals as well as there is no modification to the contents of the navigation message itself. Now, the attacker begins to manipulate the spoofing signal such that the receiver computes a false location and begins to alter its course. The attacker can either slowly introduce a temporal shift from the legitimate signals or directly manipulate the navigation message contents to slowly deviate the course of the ship to a hostile destination. Tippenhauer et al.~\cite{Tippenhauer2011} describe the requirements for an attacker to execute a seamless takeover and move the target receiver towards the intended location.

%

\subsection{Performance of Existing Countermeasures}
\label{subsec:proposed_countermeasures}
In this section, we discuss existing countermeasures and describe their effectiveness against  various types of spoofing attacks. A number of countermeasures were based on detecting anamolies in the physical-layer characteristics of the received signal. In addition to the estimated position, velocity and time, modern GPS receivers output information pertaining to certain physical-layer characteristics directly as \emph{receiver observables}. Modern GPS receivers can be configured to output, for example, automatic gain control (AGC) values, received signal strength (RSS) from individual satellites, carrier phase values, estimated noise floor levels \etc A number of previous works~\cite{Akos2012,Bastide2001,Warner2003} proposed using some of the above mentioned receiver observables to realize spoofing awareness in a GPS receiver. For example, in~\cite{Warner2003} the authors suggest monitoring the absolute and relative signal strength of the received satellite signals for anomalies, number of visible satellites (should not be high), simultaneous acquisition of satellite signals, \etc Other countermeasures such as detecting sudden changes to the AGC values were also proposed for detecting GPS spoofing attacks. Automatic Gain Controller (AGC) is a hardware module that varies the gain of the internal amplifier depending on the strength of the received signal. Such countermeasure are at best capable of detecting attackers who transmit their spoofing signal at very high power. They are ineffective against attackers who have better control over their spoofing signal.

Several spoofing detection strategies based on analyzing the distortions present in the output of the receiver's correlation function were proposed in~\cite{Wesson2011,phelts2001multicorrelator}. In an ideal noise-free environment, the correlation output has minimal distortions. The authors argue that during a spoofing attack, the attacker's signal would distort the output of the correlators, which can be used to detect the attack itself. However, the correlation output is also distorted due to multipath signals that arrive a few nanoseconds later than the direct signal. Wesson et al.~\cite{Wesson2011} showed that it is indeed difficult to distinguish between the distortions caused due to a spoofing attack and a legitimate multipath signal. Spoofing detection techniques based on the differences in the inherent spatial characteristics of the received signal such as direction or angle of arrival~\cite{Montgomery2009,Psiaki2011,broumandan2012gnss} also face the same challenge of reliably distinguishing between legitimate multipath signals and a spoofing attack. Additionally, they also require additional hardware modifications to the GPS receiver. To summarize, although several countermeasures have been proposed in literature to detect spoofing attacks, there is no countermeasure today that is effective in detecting strong attackers such as a seamless takeover attack. Moreoever, there is no platform that can be used to compare and evaluate the effectiveness of existing countermeasures in real-world scenarios. Today, it is still possible to spoof a victim receiver to any arbitrary location without being detected.

\section{\rxname~-- A Spoofing Resilient GPS Receiver}
\label{sec:spoofing-aware-gps-receiver}

The design of \rxname is largely motivated by the lack of a GPS receiver capable of detecting or constraining all the spoofing attacks known in literature. In this section, we present the design of \rxname, the first GPS receiver capable of detecting or constraining all known spoofing attacks. Our receiver design consists of two key components: (i) Auxiliary Peak Tracker (APT) and (ii) Navigation Message Inspector (NAVI) module. First, we describe the auxiliary peak tracking module, a novel countermeasure which plays a vital role in constraining even a strong attacker capable of a seamless takeover. The key feature of APT is that it acquires and tracks not only the strongest received satellite signal but also the weaker signals that may be present in the environment. Second, we introduce a navigation message inspector (NAVI) which inspects the decoded contents of the navigation message from every satellite and reports any discrepancies. We show that NAVI is capable of detecting attackers who modify the contents of the navigation message. The Auxiliary Peak Tracker protects \rxname from attackers who are not synchronized (non-coherent) to the legitimate GPS signals currently being received and the Navigation Message Inspector prevents attackers from modifying the contents of the navigation message. The combination of auxiliary peak tracking and the navigation message inspector enables \rxname to reliably detect all types of spoofing attacks. 

\subsection{Auxiliary Peak Tracking (APT)}
In this section, we describe the details of our proposed Auxiliary Peak Tracking technique, which is one of \rxname's key features that makes it resilient to spoofing attacks. Typically, GPS receivers have multiple acquisition and tracking modules to simultaneously search and track different satellites. Each set of acquisition and tracking module is called a “channel” and each satellite signal is acquired and tracked by only one channel. For example, a $24$-channel GPS receiver can simultaneously search for $24$ satellites thereby shortening the time to acquire a position fix when compared to a $4$-channel receiver. In other words, the receiver searches for a satellite by allocating each channel to one specific satellite. The receiver acquires or searches for a particular satellite signal by correlating its own replica of that specific satellite's pseudorandom code with the received signal. If the search results in a correlation value above a certain threshold, the receiver then switches to tracking and demodulating the navigation message data. \textit{It is important to note that GPS receivers acquire and track only the satellite signal that produces the strongest correlation peak and ignore any weaker correlation peaks as noise}.

\begin{figure}[t]
  \centering
  \includegraphics[width=\columnwidth]{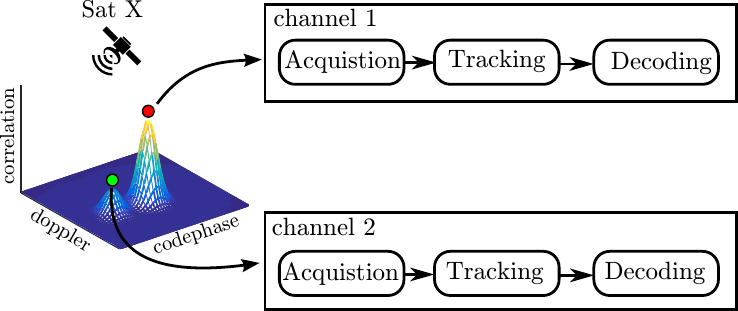}
  \caption{Auxiliary Peak Tracking (APT): \rxname uses more than one channel to acquire, track and decode each satellite's signal. This enables tracking of signals that produce weaker acquisition correlation peaks.}
  \label{fig:spoof-aware-receiver-design}
\end{figure}

In SPREE, we allocate \textit{more than one} channel to the same satellite. This means that in addition to tracking the signal that results in the strongest correlation, \rxname can also track weaker correlation peaks (if pres\-ent) for the same satellite. In other words, \rxname does not restrict itself to the satellite signals that produces the maximum correlation, but it also detects and tracks signals that produces weaker correlation (Figure~\ref{fig:spoof-aware-receiver-design}).\\

\noindent\textbf{Spoofing detection by tracking auxiliary peaks:} The Auxiliary Peak Tracker protects \rxname from attackers who are not synchronized to the authentic GPS signals. Recall that the attacker transmits spoofing signals with a higher power such that the authentic GPS signals are overshadowed. Even though the spoofing signals have successfully overshadowed the authentic signals, they are still present in the environment and it is difficult for an attacker to completely annihilate them. In order to completely annihilate authentic GPS signals, the attacker first needs to know the precise location (cm level) of the receiver. Furthermore, he needs to annihilate all the multipath components of the GPS signal at the receiver. This means that the attacker should be able to transmit nulling signals such that they cancel both the direct GPS signal and \textit{all} the possible multipath components at the receiver. In case the receiver is in motion, the attacker must be able to predict the \textit{exact} trajectory of the receiver. Given the difficulty of completely annihilating authentic satellite signals, they will appear as auxiliary peaks when the attacker's spoofing signals are non-coherent or in other words not synchronized with the authentic satellite signals. We provide a more detailed analysis on how \rxname's APT module enables detection of even the strong seamless takeover attackers in Section~\ref{sec:evaluation}.

\subsection{Navigation Message Inspector (NAVI)}
\label{sec:nav-message-inspection}

The Navigation Message Inspector module inspects the decoded navigation data for consistency and sanity and is key to protecting the GPS receiver from attackers who modify the contents of the navigation message. \\

\begin{figure}[t]
\centering
  \includegraphics[width=0.85\columnwidth]{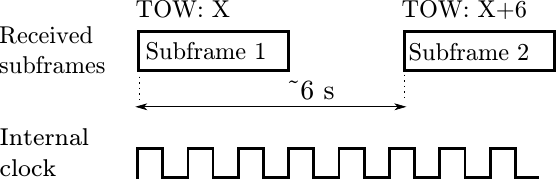}
  \caption{\rxname compares the received TOW to its internal clock and validates whether TOW is increased in 6 s intervals.}
  \label{fig:tow_clock}
\end{figure}

\noindent\textbf{Time of Week (TOW) and Receiver's Clock:}  One of the key parameters that an attacker can modify in-order to spoof a target receiver's location or time is the transmission time of the navigation messages. The navigation data transmitted by each of the satellites are divided into 5 subframes. Each subframe begins with a handover word which contains a truncated version of the time of week (TOW) at which the satellite transmitted that particular subframe. Each subframe lasts for about 6 seconds and since the TOW is transmitted once every subframe, it can only increase in steps of 6 seconds. We leverage the internal clock of \rxname's hardware and the fact that the TOW can only change in steps of $6\unit{s}$ to detect spoofing attacks (Figure~\ref{fig:tow_clock}). \rxname records the received GPS week and time of week with its internal clock count and raises an alarm if the difference in the time elapsed internally doesn't match the newly received GPS time of week. \\

\noindent\textbf{Satellite Orbital Positions:} In addition to the transmission time of the navigation message, an attacker can also modify the satellite's position in the orbit. The GPS receiver estimates the satellite's position from the ephemeris data. For example, Nighswander et al~\cite{NighswanderCCS2012} demonstrated that it is possible to modify the ephemeris data such that the receiver estimates the satellite to be in the middle of the earth. The authors executed such an attack by setting the square root of the semi-major axis of the satellite’s orbit to $0$. In our design, an attacker cannot execute such manipulations as \rxname continuously monitors and evaluates any changes to the orbital parameters.\\

\noindent\textbf{Almanac \& Ephemeris Data:} \rxname continuously monitors the decoded navigation data from all the visible satellites and performs a number of consistency checks. The almanac and ionospheric model data should be the same across all the navigation frames received from all the satellites. In addition, whenever feasible \rxname leverages the availability of navigation data such as ephemeris, almanac and the ionospheric models from third-party sources to compare the data decoded by the GPS receiver. This data is then compared against the information received from the satellites and is used to detect spoofing attacks.\\

Thus, \rxname 's navigation message inspector independently protects the receiver from attackers capable of modifying the navigation message. By combining the NAVI and APT modules, \rxname detects or constraints all types of  attacks capable of spoofing the receiver's location and time.

\section{Implementation}
\label{sec:implementation}

We implemented \rxname based on GNSS-SDR~\cite{GNSS-SDR11}, an open source software defined GPS receiver. GNSS-SDR is written in C++ and can be configured to process signals received directly from a radio hardware platform such as USRP~\cite{ettus_research} or from a file source. GNSS-SDR works with a range of hardware platforms and signal recorders such as USRP, SiGe GN3S Sampler, NSL Primo~\cite{nsl_primo}, IFEN's NavPort~\cite{ifen} etc. The architecture of GNSS-SDR largely resembles the design of a typical GPS receiver as described in Section~\ref{sec:gps-overview}. It consists of a signal source and a conditioner module which are responsible for interfacing with the underlying receiver hardware or file source. Similar to typical GPS receivers, GNSS-SDR also consists of several \textit{channels}; each individual channel managing all the signal processing related to a single satellite. In GNSS-SDR, the \textit{channel} is a software module that encapsulates the functions of acquisition, tracking and navigation message decoding blocks. All the channels then report to a module that estimates the pseudoranges and a number of other observables. Finally, if enough information is available, the receiver calculates a position, velocity and time. A configuration file allows the user to chose operational parameters such as the sampling frequency, the algorithms to use for each processing block, signal source \etc We modified the acquisition and tracking modules of GNSS-SDR to realize \rxname. First, we implement the auxiliary peak tracking system within the GPS receiver’s acquisition module. Recall that the auxiliary peak tracker enables the receiver to track multiple signals of the same satellite instead of limiting it to the strongest component only. We implement the navigation message inspector which checks the consistency and sanity of the extracted navigation data within the tracking module of the receiver.\\ 

\noindent\textbf{Auxiliary Peak Tracking (APT):} In SPREE, when a particular satellite is assigned to a channel, all local peaks of the acquisition correlation function, which are above a certain threshold are collected and stored for processing. This is in contrast to the modern receivers only choosing the highest correlation peak. Each local peak is then assigned to a different channel in descending order of magnitude for tracking. The maximum number of channels that can track the same satellite is made configurable at run time. The number of channels that can be assigned to track the same satellite will influence the number of peaks that can be evaluated at the same time.

If \rxname is successful in acquiring more than one peak, it records the differences in their arrival times \ie the separation between two peaks.  If the difference is more than the maximum acceptable time difference, $\tau_{max}$, \rxname detects a spoofing attack. The value  $\tau_{max}$  is set in the configuration file. This check is done each time a new navigational message is received. The arrival time is computed in the tracking module, where it is estimated based on the sample counter of GNSS-SDR and fine tuned based on the code phase of the satellite signal. After an auxiliary peak has been acquired, tracked and evaluated for signs of spoofing and none are found it is dropped and the channel is free to acquire another auxiliary peak to evaluate. If the peak remains it will be evaluated again when a channel is free \textit{and} all other peaks have been evaluated.\\


%

\noindent\textbf{Navigation Message Inspector (NAVI):} In GNSS-SDR, a telemetry decoder is responsible for decoding the contents of the received navigational message. First, \rxname records the time of week decoded from each of the received navigation message subframes. If the  difference in time of week present in consecutive subframes does not match with its internal clock count (more than 6 s difference due to the minimum resolution), \rxname raises an alarm. Next, the stored navigation data for each of the visible satellites is compared with the contents of the preceding navigation message for that particular satellite. If there is a discrepancy between these two values, \rxname notes it as a possible spoofing attack. Also, \rxname compares the navigational data from all satellites with each other for any discrepancies in the almanac and ephemeris data. Recall that, the almanac and ionospheric model data should be the same across all the navigation frames received from all the satellites. If configured to do so and if possible, it can also compare the time, almanac, ephemeris and the ionospheric model data received from the satellites to data received from third-party sources using the Secure User Plane Location (SUPL) protocol. These checks are done each time a new navigation message is received.\\

In addition to the above modules, we also implement several existing countermeasures described in Section~\ref{sec:gps-spoofing-attack} to facilitate real-world performance evaluations. However, we restrict our discussion to our main contributions, the APT and NAVI module as they enable reliable detection of all known spoofing attacks in literature. It is important to note that~\rxname adds no additional requirements on the underlying hardware and supports all the platform and file sources supported by GNSS-SDR.

\section{Security Evaluation}
\label{sec:evaluation}
\begin{figure}[t]
\centering
  \includegraphics[width=0.7\columnwidth]{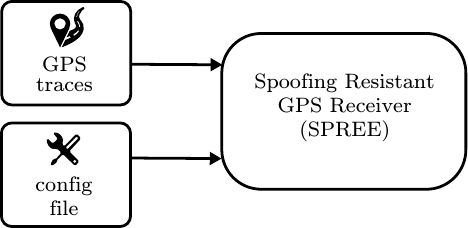}
  \caption{Evaluation Setup: A configuration file specified vital system parameters such as input source, source signal sampling rate and configuration of the spoofing detection module.}
  \label{fig:evaluation-setup}
\end{figure}
In this section, we evaluate \rxname and present its security guarantees. Figure~\ref{fig:evaluation-setup} shows our evaluation setup. A configuration file is used to select \rxname's parameters including those needed by the spoofing detection module. In our evaluations, the GPS signal traces (spoofing and clean) were recorded and stored in files and later input to \rxname. We evaluated \rxname against three different sets of GPS signals: (i) a public repository of spoofing traces (TEXBAT)~\cite{humphreys2012texas}, (ii) signals recorded through our own wardriving effort and (iii) spoofing signals generated using COTS GPS simulators.\\


\subsection{GPS Traces}
\noindent\textbf{GPS Simulator:} First, we evaluated the performance of \rxname against our own spoofing signals generated using commercially available GPS simulators. Specifically, we used Spectracom's GSG-5 Series advanced GPS simulator~\cite{spectracom} in order to generate our spoofing signals. One of the key features of the simulator is its ability to generate multipath signals for any satellite. It is even possible to configure the multipath's power levels and time offset \ie the extra distance travelled by the multipath relative to the original line-of-sight (LOS) signal. The GPS simulator traces were mainly used to evaluate the ability of \rxname to robustly detect auxiliary peaks. In addition, we used the GPS simulator traces to simulate attackers capable of manipulating the content of the navigation messages.\\

\noindent\textbf{Texas Spoofing Test Battery (TEXBAT):} TEXBAT~\cite{humphreys2012texas} is a set of digital recordings containing GPS spoofing tests conducted by the University of Texas at Austin. TEXBAT is the only publicly available dataset and the de-facto standard for testing spoofing resilience of GPS receivers. TEXBAT includes two clean (spoofing free) data sets in addition to spoofing scenarios based on the location and time of the clean GPS traces. The set of spoofing traces contain a wide variety of scenarios including take-over attacks where either the time or position of the target receiver is spoofed. The spoofing signals are closely code-phase aligned with the authentic signals. Furthermore, the carrier phase of the seamless takeover scenarios are aligned with the authentic signals during the takeover.\\

\begin{figure}[t]
  \centering
  \includegraphics[width=0.85\columnwidth]{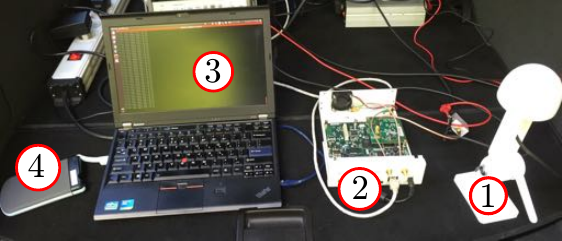}
  \caption{Our wardriving setup with a front-end consisting of a (1) a active conical GPS antenna and a (2) USRP N210. The signals were recorded using a (3) laptop. The recording were periodically moved to an (4) external hard disk.}
  \label{fig:wardriving}
\end{figure}

\noindent\textbf{Wardriving:} In addition to using TEXBAT scenarios, we collected our own  GPS traces through an extensive wardriving effort. We used the wardriving dataset to evaluate \rxname's behaviour in a non-adversarial (only legitimate GPS signals present) scenario and determine how reliable is \rxname with respect to false alarms. The setup used for recording the GPS signals during the wardriving effort is shown in~\ref{fig:wardriving}. The front end of the setup consists of an active conical GPS antenna and a bias-tee. We used an USRP N210 and GNURadio and recorded raw GPS signals into an external hard disk. The signals were sampled at $10\unit{MHz}$ and stored in complex data format. The setup itself was powered through the car's power outlet. We recorded the GPS signals at various locations: (i) An open field, (ii) parking lot of a small village, (iii) driving on a highway, (iv) driving inside a city, (v) inside a city with neighbouring tall buildings and (vi) inside a forest with dense tree cover. 

\subsection{Security Evaluation}
Recall that an attacker can influence the receiver's estimates by either manipulating the contents of the navigation messages or temporally shifting the navigation message signals while transmitting the spoofing signals.\\
\begin{figure}[t]
\centering
  \includegraphics[width=0.85\columnwidth]{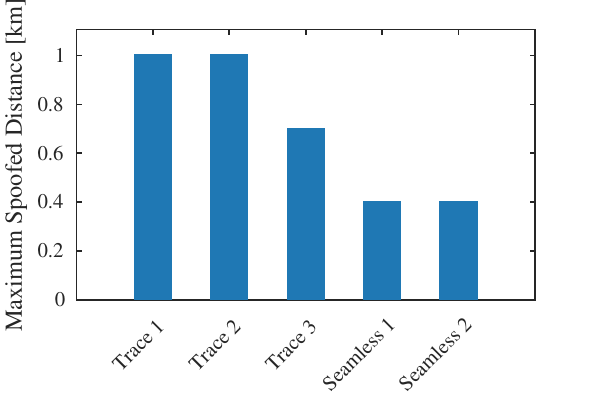}
  \caption{Spoofing detection in TEXBAT dataset. \rxname detected auxiliary peaks in all the spoofing traces. The maximum location offset the attacker could cause before being detected was less than a kilometer.}
  \label{fig:texbat_auxpeaks}
\end{figure}

\begin{figure*}[t]
  \centering
  \includegraphics[width=0.9\textwidth]{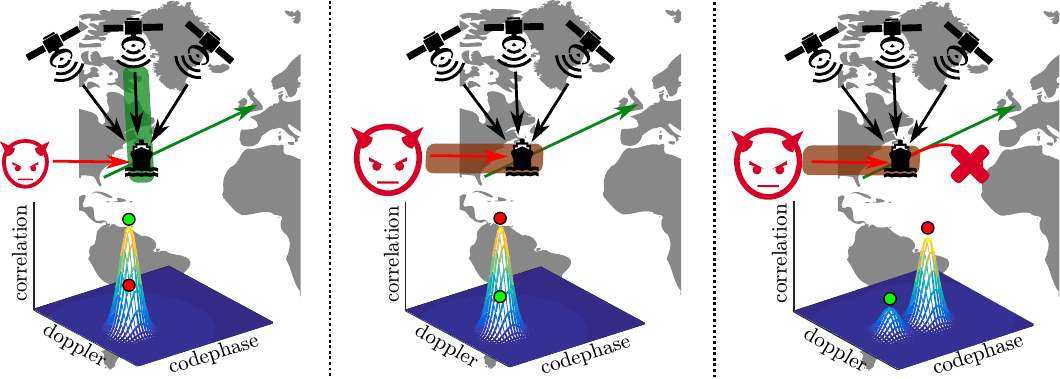}
  \caption{Detecting seamless take over attack. As the attacker begins to drift the spoofing signal away with the intention of changing the course of the ship, SPREE will detect the auxiliary peak produced by the legitimate satellite signal and rise an alarm.}
  \label{fig:seamless-takeover-detection}
\end{figure*}
\noindent\textbf{Detecting Non-coherent Attackers:} Recall that a non-coherent attacker's spoofing signal is not synchronized with the authentic satellite signals. Even though the receiver might be locked on to the attacker's spoofing signals, the authentic signals will appear as auxiliary peaks due to the Auxiliary Peak Tracking module. The effectiveness of detecting such non-coherent spoofing attacks depends on the ability of the APT module to detect and track auxiliary peaks. First, using our own GPS simulator traces, we tested the ability of the APT module to detect and track multiple acquisition correlation peaks. Specifically, we leveraged the ability of the simulator to generate duplicate copies of a satellite signal at different time intervals away from the original signal. We generated signal copies spaced between $50\unit{ns}$ to $1000\unit{ns}$ and determined that our receiver was able to reliably detect and track auxiliary peaks spaced $500\unit{ns}$ or more. In some scenarios, it was able to track peaks much closer, however not reliably (over multiple runs). Thus, we configured APT module to track auxiliary peaks that are separated by more than $500\unit{ns}$. The choice of $500\unit{ns}$ separation between two peaks for spoofing detection is supported by two additional reasons: (i) During signal acquisition (searching for satellite signals), GPS receivers shift their correlator typically by half a chip\footnote{A chip is one bit of the pseudorandom code} period \ie $500\unit{ns}$. This means that most modern receivers can reliably track peaks that are separated by $500\unit{ns}$ and no additional hardware changes are required to implement \rxname in modern receivers. (ii) Several prior works on modelling GNSS multipath signals~\cite{kong2011statistical,Braasch2001,lehner2005land,manandhar2006gps} show that most GPS multipaths are delayed by less than $300-400\unit{ns}$. This means that it is highly unlikely to observe an auxiliary peak caused due to legitimate multipath signals occurring at more than $500\unit{ns}$ away from the line-of-sight signal peak. Moreover, the attenuation and polarization shift introduced in the legitimate signals due to reflections that are a few hundred metres away would make the signal untrackable. We proceeded to evaluate \rxname against the TEXBAT set of GPS spoofing signal traces described previously. \rxname detected auxiliary peaks in all the traces containing spoofing signals and failed to detect any auxiliary peaks for the clean non-spoofing traces. Based on the separation of auxiliary peaks at the time of detection, we evaluated the maximum possible location offset an attacker could have caused without being detected and present it in Figure~\ref{fig:texbat_auxpeaks}. In the case of the seamless takeover attacks, the maximum deviation an attacker could introduce in \rxname was about 400 m. It is important to note that traces 1, 2 and 3 contain spoofing signals that are not as closely synced as the seamless takeover traces and hence the larger values for maximum spoofed distance. For completeness, we processed our wardriving traces that represent clean, non-spoofing scenarios for any false alarms. \rxname did not detect any auxiliary peaks.\\


\noindent\textbf{Detecting Navigation Message Modifications:} We will now analyze \rxname's resilience against attackers who modify the contents of the navigation message. The key parameters that an attacker can manipulate in the navigation data are the time of transmission and the satellite's orbital information present in the almanac and ephemeris.\\ 

\noindent\textit{Modifying TOW:} As described in Section~\ref{sec:nav-message-inspection}, the value of TOW can be altered only in steps of 6 seconds. \rxname leverages the internal clock of the hardware receiver to continuously compare the received TOW data against its internal clock count. \rxname raises an alarm if the difference in the time elapsed internally doesn't match the newly received GPS time of week information. We note that even a watch crystal today has an error rating of approximately 10 ppm which is a drift of less than a second in one day. Therefore a drift of $6\unit{s}$ can be easily detected even without a thermally controlled crystal oscillators (TCXO\footnote{Modern TCXOs have error ratings between $1-100\unit{ppb}$ and are available for under \$10}) that is present in modern hardware receiver platforms. We evaluated \rxname against such an attack using two GPS simulators each spoofing the same satellite however with different TOW data and \rxname successfully detected the attack. Both the simulators were synchronized to the same reference clock signal. We used this setup to evaluate \rxname's resilience to attacks described in~\cite{NighswanderCCS2012} such as arbitrary manipulation of week numbers and date desynchronization attacks.\\

\noindent\textit{Modifying Ephemeris Data:} The attacker can also manipulate the ephemeris data to force the receiver to malfunction. Ephemeris data gets updated once every two hours and contain precise satellite orbital information including satellite clock biases. However, it was shown in~\cite{NighswanderCCS2012} that it is trivial to force a receiver to accept ephemeris changes whenever possible. Since \rxname's NAVI module keeps track of the elapsed time using the receiver's internal clock, it can be configured to ignore any ephemeris updates within the 2-hour time interval. It is also important to note that any changes to the satellite orbital information or in general the ephemeris data can be compared against ephemeris data available from third-party sources~\cite{nasa_dailyephemeris}. Additionally, \rxname is capable of recording the ephemeris data received from all satellites in the past and notify if there is any unexpected change in the ephemeris data values.\\

\noindent\textbf{Detecting Seamless Takeover Attack:} As described previously, a seamless takeover attack is an attack in which the attacker takes control of the victim receiver without any disruption to its current state. This type of attacker is one of the strongest attackers known in literature and no existing countermeasure is effective in detecting the seamless takeover attack. We will now see how \rxname enables detecting a seamless takeover attack. Consider the same example of a ship on its way from United States to the UK, currently locked on to legitimate satellite signals. The attacker begins a seamless takeover by transmitting spoofing signals that are synced to the legitimate satellite signals but at a lower power level. The output of the acquisition module is shown in Figure~\ref{fig:seamless-takeover-detection}. Notice that the legitimate satellite signal (shown in green) is stronger than the spoofing signal but are synchronised to each other. Now the attacker, increases the spoofing signal’s power and takes over the receiver. Note that, even though the receiver is locked on to the attacker, there is still no change in route yet. This is because the attacker is both synchronised to the legitimate signals and is transmitting the same navigation message. Now, the attacker begins to drift the spoofing signal away with the intention of changing the course of the ship. At this point, a typical GPS receiver will ignore any weaker correlation peaks that exist and compute its location based on the attacker’s signal. However, SPREE will detect an auxiliary peak and rise an alarm.\\

\begin{figure}[t]
\centering
  \includegraphics[width=0.85\columnwidth]{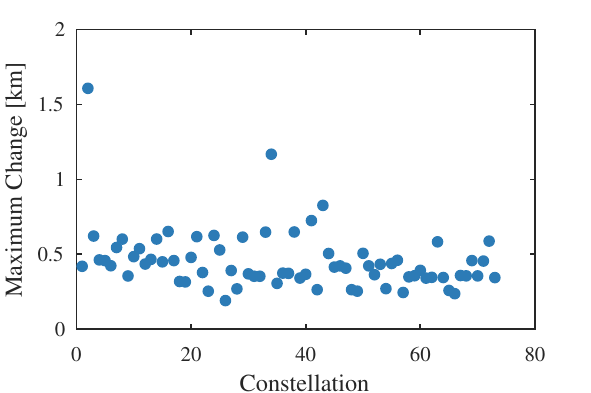}
  \caption{Maximum location offset. An analysis of 73 satellite constellations (as observed during wardriving) show that a strong attacker can cause a maximum location offset of less than 1 km in majority of the scenarios before being detected.}
  \label{fig:constellation}
\end{figure}

\noindent\textbf{Maximum position offset:} Recall that, \rxname detects any modifications to the contents of the navigation message and tracks peaks of the same satellite that are separated by more than $500\unit{ns}$. This value was setup after extensive experiments using signals from GPS simulators and our own wardriving efforts as described previously. This means that the attacker is limited to temporally shifting his spoofing signals by at most $500\unit{ns}$ which results in a $150\unit{m}$ change in the pseudorange estimated by the receiver for that specific satellite. It is important to note that the effect of this change in pseudorange caused by the attacker on the receiver's final position estimate depends on the constellation of the satellites. We collected all the different constellations observed during our wardriving and evaluated the effect of temporally shifting the satellite pseudoranges by $150\unit{m}$. Our analysis accounted for all possible pseudorange changes an attacker can introduce on all combinations of visible satellites. We analyzed over $73$ different satellite constellations, each one with four satellites, and calculated the maximum possible location offset an attacker could introduce. Our results are shown in Figure~\ref{fig:constellation}. On an average the maximum position deviation was $455\unit{m}$. This means that \eg in the ship hijack scenario, it would not be possible for an attacker to deviate the course of the ship by more than $455\unit{m}$. Note that, we limited our analysis to constellations consisting of only four visible satellites, which is the most favourable for an attacker. In most environments, more than four satellites will be visible, which will further constrain how much the attacker can change the victim's position. Furthermore, we observed that the constellations that allow the attacker to spoof the receiver more than 1~km away, comprised satellites at very low elevation angles. Therefore, configuring \rxname to only accept satellite signals with a minimum elevation angle will potentially constrain the attacker further.\\

\section{Discussion}
\label{sec:gps_discussion}
\noindent\textbf{Integrating \rxname into commercial receivers:} One of the main differences between \rxname and a commercial GPS receiver is that unlike commercial receivers which track one satellite per channel, \rxname uses multiple channels to track the same satellite. This means that without any hardware changes \ie for the same number of acquisition and tracking channels, our spoofing-aware receiver will track less number of satellites than its capable of. In order to do this, two changes are necessary: (i) allocate a minimum of two channels for every visible satellite signal (one for the authentic GPS signal and one that keeps searching for a potential spoofing signal) and (ii) search the entire range of time delays for weaker acquisition peaks. The number of channels allocated per visible satellite signal can be easily modified in the firmware. However, as mentioned before, this will limit the number of satellites that the receiver can simultaneously acquire and track. Modern receivers typically have $32-128$ channels capable of tracking $32-128$ satellites simultaneously\footnote{sometimes used in receiver's capable of using more than one satellite navigation system such as GLONASS} and allocating two channels for each satellite will reduce the number of satellites that can be tracked by half. In reality, this is not a problem since the typical number of visible satellites at any time instant is not more than 10 or 11. In order to track auxiliary peaks, we keep a list of all auxiliary peaks found during the acquisition in a float array. The number of floats stored is $2 \cdot \frac{F_s}{1000}$ for each acquisition, where $F_s$ denotes the sampling rate. This means that for a sampling rate of 10 MHz each acquisition requires an additional $\approx 19.5\unit{kB}$ of storage. We believe this to be negligible when compared to the available RAM in most of the modern receivers today. There is practically no performance overhead in detecting changes to the contents of the navigation message by an attacker. The only waiting time is the time ($\approx 6 \unit{s}$) needed to receive and decode the new subframe completely. Hence our design modifications can be easily integrated into a modern GPS receiver with only a firmware upgrade and does not require any changes to the underlying hardware.\\

\noindent\textbf{Probability of False Alarms:} False alarms can be caused due to an event that forced \rxname to believe it is being spoofed. In the case of the auxiliary peak tracking module, the arrival of a legitimate multipath signal with a delay of more than $500\unit{ns}$ and with a signal strength greater than the acquisition threshold will result in \rxname raising a spoofing alert. This is unlikely to be captured by the GPS receiver due to the following reasons: (i) change in polarization--GPS signals are typically right hand polarized and any reflections causes a change in the polarization of the signal. Majority of GPS receiver antennas are configured to received the direct right hand circularly polarized signals and attenuate reflected signals. (ii) Propagation path loss--Since the multipath signals travel a few hundred metres more than the direct line of sight signal, the signals undergo more attenuation due to propagation path loss. Also, reflections from surfaces themselves may cause the GPS signal to attenuate and therefore, given the received power levels of direct line of sight GPS signals on the ground, multiple reflections would eventually only make the signal untrackable. In addition, auxiliary peaks caused by legitimate multipaths tend to be momentary and untrackable in contrast to a peak caused by a seamless takeover attack. Recall that, \rxname did not detect any auxiliary peak beyond the set $\tau_{max}$ of $500\unit{ns}$ on the traces collected during our wardriving effort. In fact, an analysis of the temporal behaviour of multipath signals against spoofing signals can enable distinct identification of peaks caused due to a spoofing signal. Note that, even after detecting auxiliary peaks, it is currently difficult to distinctly identify the peak caused by the spoofing signal and that caused by the legitimate signal. Thus the results of the temporal behavioural analysis can help the receiver to ignore or internally cancel the spoofing signal and thereby building better resilience to spoofing attacks.

\section{Related Work}
\label{sec:related-work}

The work that comes closest to \rxname is the design of an inline anti-spoofing device~\cite{ledvina2001line}. The device connects between the GPS antenna and a GPS receiver and uses complex correlation peak distortion techniques to identify spoofing signals. As demonstrated in~\cite{Wesson2011}, such countermeasures face the challenge of distinguishing spoofing signals from real-world channel effects and are ineffective against seamless takeover attackers. Also, the device is incapable of detecting attackers who modify the contents of the navigation messages. Several works~\cite{kuhn2005asymmetric,lo2010authenticating,wesson2012practical} propose solutions that are cryptographic in nature and therefore require modifications to the GPS infrastructure. Incorporating cryptographic authentication into civilian GPS, similar to military GPS, could to an extent mitigate spoofing attacks. However this would require distribution and management of shared secrets which makes it infeasible for a large set of applications. Additionally, cryptographic authentication does not protect against signal replay attacks where an attacker simply records legitimate GPS signals at one location and replays it to the victim receiver~\cite{Papadimitratos2008}. Some other proposals depended on additional hardware such as additional receivers, alternative navigation systems, sensors etc. Tippenhauer et al.~\cite{Tippenhauer2011} proposed the use of multiple synchronized GPS receivers to detect spoofing. They show that spoofing a set of synchronized GPS receivers, with known relative distances or geometrical constellation restricts the number of locations from
where an attacker can transmit the spoofing signals. Cross-validation of the position estimates against alternate navigation systems such as Galileo~\cite{hofmann2007gnss} were also proposed. However, a simulator that can spoof both GPS and Galileo will easily defeat this countermeasure. Data from other sensors can also be used to cross validate GPS navigation solutions. For example, inertial measurement units (\eg accelerometer,
gyroscope, compass) have already been proposed as alternative ways to navigate
during temporary GPS outages~\cite{titterton2004strapdown,farrell1999global,wendel2006integrated}.
The main drawback of inertial navigation units is the accumulating error of the
sensor measurements. These accumulated sensor measurement errors affect the
estimated position and velocity over longer duration of time and hence limit
the maximum time an IMU can act independently.

\section{Conclusion}
\label{sec:gps_conclusion}
In this paper, we presented \rxname, the first GPS receiver that detects all known spoofing attacks. We designed, implemented and evaluated \rxname against different sets of signal traces and showed that even a strong attacker capable of a seamless takeover cannot deviate the receiver by more than 1 km. This is a vast improvement over current GPS receivers that can be spoofed to any arbitrary location in the world. Finally, we release our implementation and the GPS dataset used in our evaluations to the research community. 

\bibliographystyle{acm}
\bibliography{gps}
\subsection*{Appendix}
\begin{table}[h]
\centering
\begin{tabular}{|c|l|}
\hline
 \textbf{Trace Set} & \textbf{Description}\\ \hline
 Trace 1 & (ds3) Static Matched Time Push\\
 Trace 2 & (ds2) Static Overpowered Time Push\\
 Trace 3 & (ds4) Static Matched Position Push\\
 Seamless 1 & (ds7) Seamless Carrier Phase Aligned\\
 Seamless 2 & (ds8) Security Code Estimation Replay \\
 \hline
\end{tabular}
\caption{TEXBAT Spoofing Traces as mapped in Figure~\ref{fig:texbat_auxpeaks}}
\label{tab:vsd_results}
\end{table}

\end{document}